\documentclass[twocolumn,twoside]{revtex4}
\usepackage{graphicx}
\usepackage{fancyhdr}
\begin{document}

\title{BESIII Exotics}

%

\author{G.~Mezzadri\\
on behalf of the BESIII Collaboration}
\affiliation{INFN - Sezione di Ferrara, Italy 44122}

\begin{abstract}
Since few years, a new family of exotic states has been appearing above the open-heavy meson thresholds: the so-called $XYZ$ states. BESIII at the BEPCII $e^+e^-$ collider plays a unique role in the study of those particles in the charmonium sector. Changing the beam energy, BESIII can collect large data samples by means of scans of the resonant region, accessing directly to all vector states. As part of a larger upgrade program, BESIII has planned to increase the center of mass energy to reach $4.7$ GeV: this will allow BESIII to investigate the nature of the $Y(4660)$, that was at first observed by Belle and BaBar after Initial State Radiation only in $\pi\pi \psi(2s)$ and $\Lambda_c$ $\bar{\Lambda}_c$ final states. The relative branching ratio seems to point toward a baryonium interpretation of the resonance, as expected in Rossi-Veneziano model. BESIII can directly measure the cross sections around the expected peak position and verify this prediction. In this presentation, the status of the $XYZ$ searches at BESIII will be presented, with a focus also on the plans for the newest data taking and for the $Y(4660)$ studies.

\end{abstract}

\maketitle

\thispagestyle{fancy}


\section{Introduction}
In the energy region above the $DD^*$ threshold a new series of structures have started to appear since 2003, when the $X(3872)$ was discovered by BELLE Collaboration \cite{belle_x3872}. These structures cannot be identified as conventional charmonia because either they are in over-abundance  with respect to the predicted one (as in the case of $1^{--}$ $Y$ states) or they have non-zero electrical charge (as in the case of $Z$ states). It exists another class of structures, the $X$ ones, that includes all the remaining states that are neither charged nor vector-like. A detailed overview on these states can be found in Ref.~\cite{olsen}.

While the number of observed states grows up, the clearness on their nature has not followed through. There are several models that can explain some properties, but there is no consensus on what is the underlying structure of this family of states. A proposal about the nature of these states is presented in these proceedings in Ref.~\cite{Voloshin}. One of the goal of the experimental side is to establish increasing information about the properties of these states and improve the knowledge about possible connections between these states. BESIII plays an unique role in the discovery of such connections.

\section{BESIII perspective in charmonium exotics searches}
\subsection{BESIII @ BEPCII}
The BESIII (Beijing Spectrometer III) detector is hosted at the Beijing Electron Positron Collider II (BEPCII) at the Institute of High Energy Physics of Beijing, PRC. It is a third generation experiment that works in the so-called $\tau-\mathrm{charm}$ energy regime, with the possibility to tune the center of mass energy between 2.0 and 4.6 GeV. The design luminosity of $\mathcal{L} = 10^{33} \, \mathrm{cm^{-2}s^{-1}}$ at the 3.77 GeV has been achieved in 2016 in a dedicated machine development run and the world's largest data sample at one single energy in $e^+e^-$ collision has been collected, at $J/\psi$ energy, during the 2019 data taking. During the past years, BESIII has collected over 12/fb of data dedicated to $XYZ$ studies between 4.18 and 4.6 GeV and more data have yet to come.

The BESIII design follows the scheme of a typical central detector optimized for flavour physics; it has a $93\%$ of $4\pi$ acceptance and it is divided in a barrel and two endcaps regions. From the interaction point to the outside it is possible to find the Beryllium beam pipe, an Helium-based Main Drift Chamber, plastic scintillators that operates as Time-of-Flight (TOF) detector, a $\mathrm{CsI(Tl)}$ electromagnetic calorimeter, a 1 Tesla superconducting magnet, and in the return yoke of the magnet, Resistive Plate Chambers operate as Muon Detector. More details can be found in Ref.~\cite{BESIII_nim}.

\subsection{BESIII unique role}
With the possibility to tune the center of mass energy around  the peak of the charmonium-like state, BESIII can measure with unprecedented precision the cross sections of very different final states and extract the different lineshape to study the nature of the vector states. Moreover, sitting exactly beneath the peak of the production cross section can accumulate very large sample to study the possible connections between the vector $Y(4220)$ and the $X$ and the $Z$ states. This last feature is unique to BESIII, since B-factories and hadronic machines have more difficulties in studying the possible relations among the different particles either in ISR processes or in $B$ decays, where the precision is lower due to the kinematic of the process itself.

\begin{figure*}[t]
\centering
\includegraphics[width=135mm]{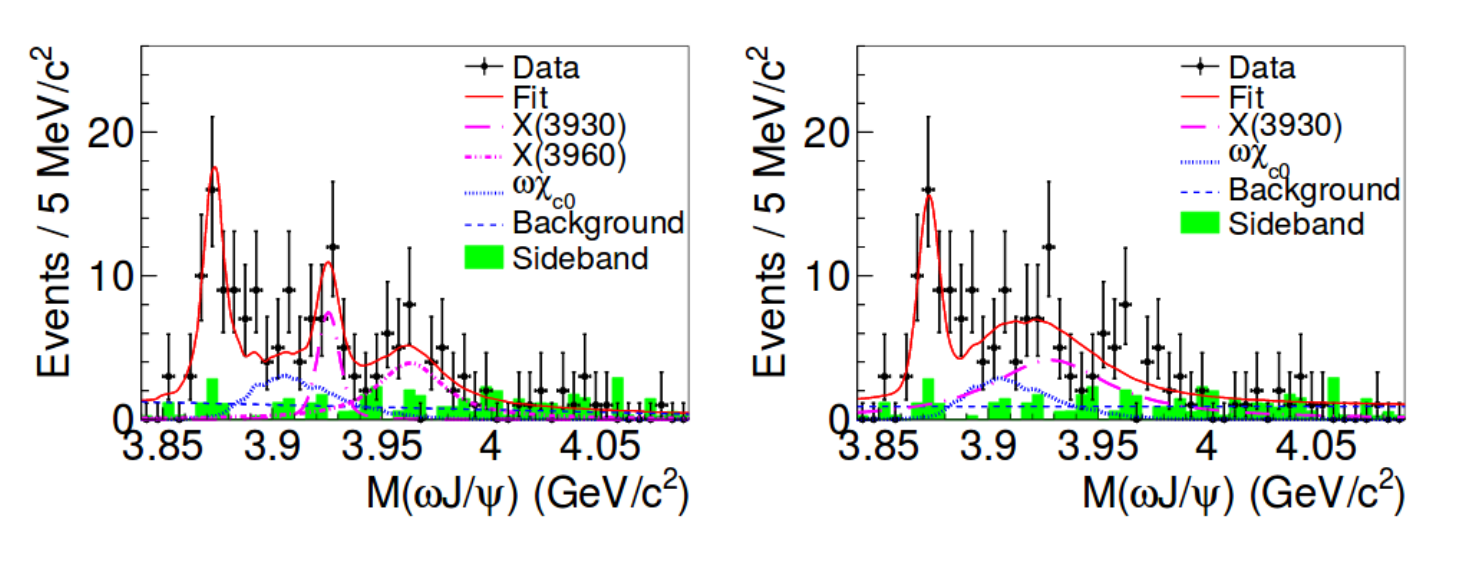}
\caption{Invariant mass of $\omega J/\psi$ in $e^+e^- \rightarrow \gamma \omega J/\psi$. The fit is performed with the $X(3872)$ plus an additional state above 3.9 GeV. The two fit hypotheses are statistically equivalent. (left) $X(3872)$ and a narrow $X(3915)$ (right) $X(3872)$ and a broad $X(3915)$.}\label{fig:figure1.pdf}
\end{figure*}

The present proceeding will show recent results based on this BESIII unique perspective.

\section{Recent Results}
\subsection{$e^+e^- \rightarrow \gamma J/\psi \omega$}

Being discovered in 2003, the $X(3872)$ is one of the most studied charmonium-like state thanks to the clear signature in the decay in $\pi\pi J/\psi$ final state. However, to address its nature it is of the utmost importance to search for many final states as it is possible. In Ref.~\cite{besiii_omegaJpsi} BESIII collaboration has reported the observation of a structure compatible with $X(3872)$ in $\omega J/\psi$ invariant mass by studying $e^+e^- \rightarrow \gamma \omega J/\psi$ with a significance of 5.1 $\sigma$. This is the first observation, while the first evidence was reported by BaBar in Ref.~\cite{babar_omegaJpsi}. In order to fit the signal, the width is fixed to the nominal value of $X(3872)$, and the mass is extracted to be $m_{X(3872)} = (3873 \pm 1.1 \pm 1.0) \, \mathrm{MeV}/c^2$. 

To fit the full spectrum above 3.9 GeV, one additional structure is needed, as shown in Fig.~\ref{fig:figure1.pdf}. Two statistically equivalent solutions are found, one narrow (left plot in Fig.~\ref{fig:figure1.pdf}) and one broad (right plot in Fig.~\ref{fig:figure1.pdf}). This structure is compatible with a $X(3915)$ observed in Ref.~\cite{gg_omegaJpsi} and in Ref.~\cite{b_to_omegaJpsi}.

\begin{figure*}[t]
\centering
\includegraphics[width=135mm]{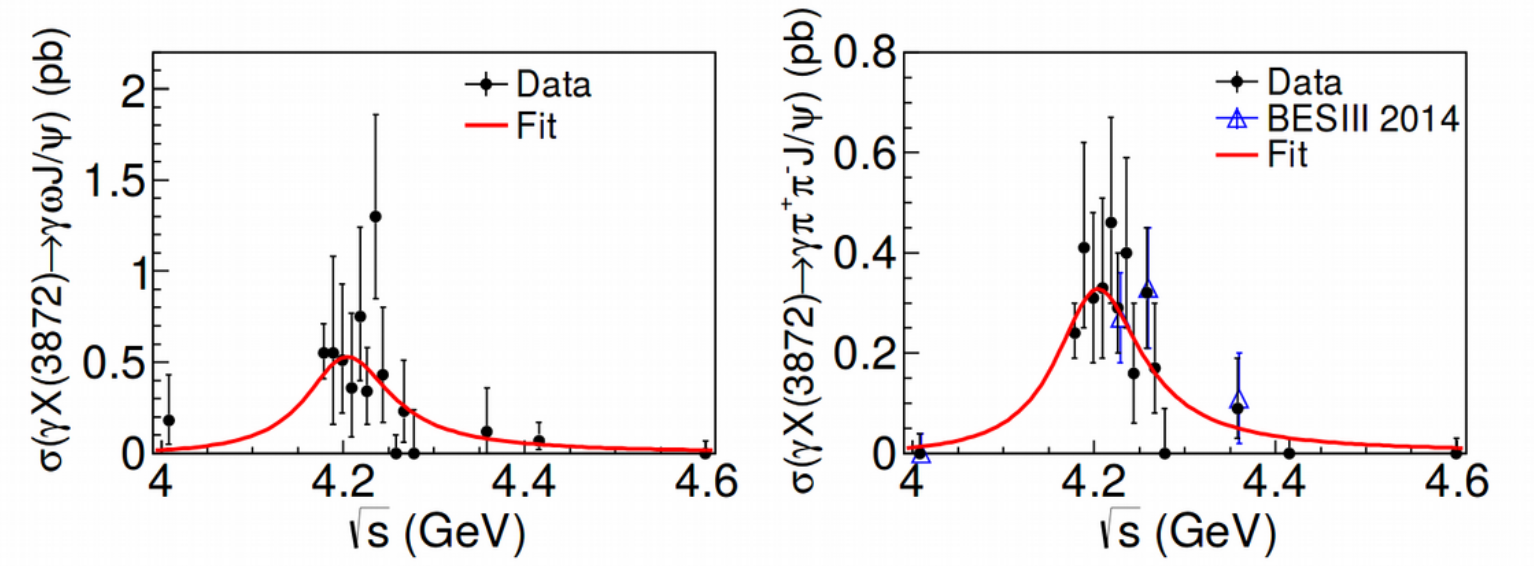}
\caption{Cross section lineshape of $e^+e^- \rightarrow \gamma \omega J/\psi$ (left plot) and $e^+e^- \rightarrow \gamma \pi \pi J/\psi$ (right plot). A simultaneous fit is performed to both lineshapes.}\label{fig:figure2.pdf}
\end{figure*}

With the possibility the process $e^+e^- \rightarrow \gamma \omega J/\psi$ it is also possible to extract the production cross section between 4. and 4.6 GeV and compare it with the production cross
section of $e^+e^- \rightarrow \gamma \pi\pi J/\psi$. This comparison is useful to address whether a transition 
$Y(4220) \rightarrow \gamma X(3872)$ 
exists and study the isospin breaking contribute in the decay of $X(3872)$. The $e^+e^- \rightarrow \gamma \omega J/\psi$ and the $e^+e^- \rightarrow \gamma \pi\pi J/\psi$ lineshapes are shown
in Fig.~\ref{fig:figure2.pdf} on the left and the right plot respectively. By performing a simultaneous fit with a Breit-Wigner it is possible to extract a structure with mass $M = (4200.6^{+7.9}_{-13.3} \pm 3) \, \mathrm{MeV}/c^2$ and width $\Gamma = (115^{+38}_{-26} \pm 12) \, \mathrm{MeV}$. This result is compatible with both the most recent $Y(4220)$ and $\psi(4150)$ parametrization. In the fit the ratio is also extracted: the result is found to be $\mathcal{R} = 1.6^{+0.4}_{-0.3} \pm 0.2$. The drop of the cross section above 4.358 GeV is not compatible with a $D\bar{D}*$ hadronic molecule calculation as proposed in Ref.~\cite{x3872_molecule}.

\subsection{Observation of $X(3872) \rightarrow \pi \chi_{c1}(1P)$}
\begin{figure}[!h]
\centering
\includegraphics[scale=0.3]{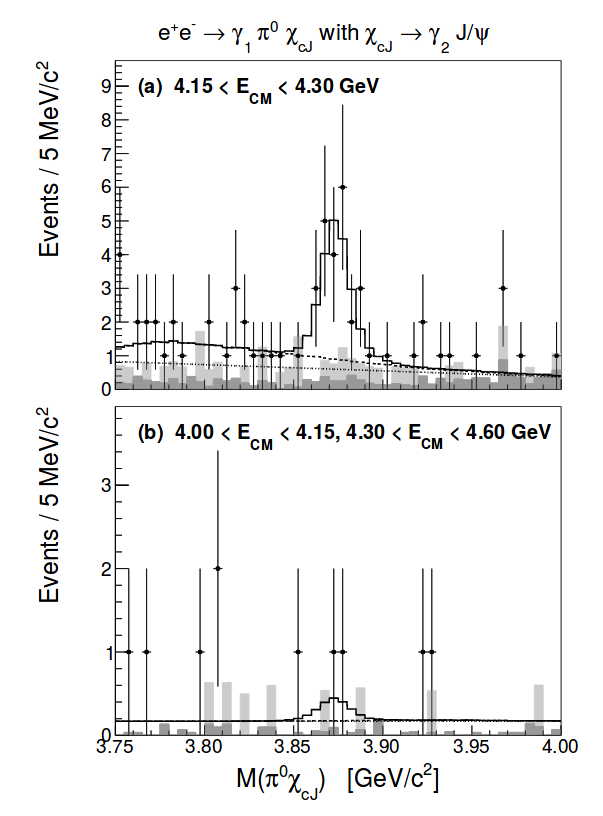}
\caption{Invariant mass of $\pi^0 \chi_{cJ}$ in $e^+e^- \rightarrow \gamma \pi^0 \chi_{cJ}$. Top plot is the invariant mass for center of mass energy in the range between 4.15 and 4.3 GeV, where the $Y(4200) \rightarrow \gamma X(3872)$ is peaking (see Fig.~\ref{fig:figure2.pdf}). Bottom plot is for other center of mass energies}\label{fig:figure3.pdf}
\end{figure}

One of the possible explanation on the nature of $X(3872)$ is that it can be a conventional $\chi_{c1}(2P)$ state. A model to test this interpretation is proposed in Ref.~\cite{voloshin_pi0}. Following this model, BESIII has reported in Ref.~\cite{BESIII_pi0} the first observation of the process $X(3872) \rightarrow \pi^0 \chi_{c1}(1P)$, with subsequent $\chi_{c1}(1P) \rightarrow \gamma J/\psi$ in $e^+e^- \rightarrow \gamma \pi^0 \chi_{c1}(1P)$. A clear signal is present in the center of mass between 4.15 and 4.3 GeV, while no signal is present for other energy values as shown in Fig.~\ref{fig:figure3.pdf}. This is compatible with the hypothesis that there is a transition between $Y(4220)$ and $X(3872)$ as already mentioned in the previous section. The present measurement highly disfavours the interpretation of $X(3872)$ as a conventional $\chi_{c1}(2P)$.

\subsection{Search for $e^+e^- \rightarrow \pi Z^{(')}_c$, $Z^{(')}_c \rightarrow \rho \eta_c$}
\begin{figure}[b]
\centering
\includegraphics[scale=0.2]{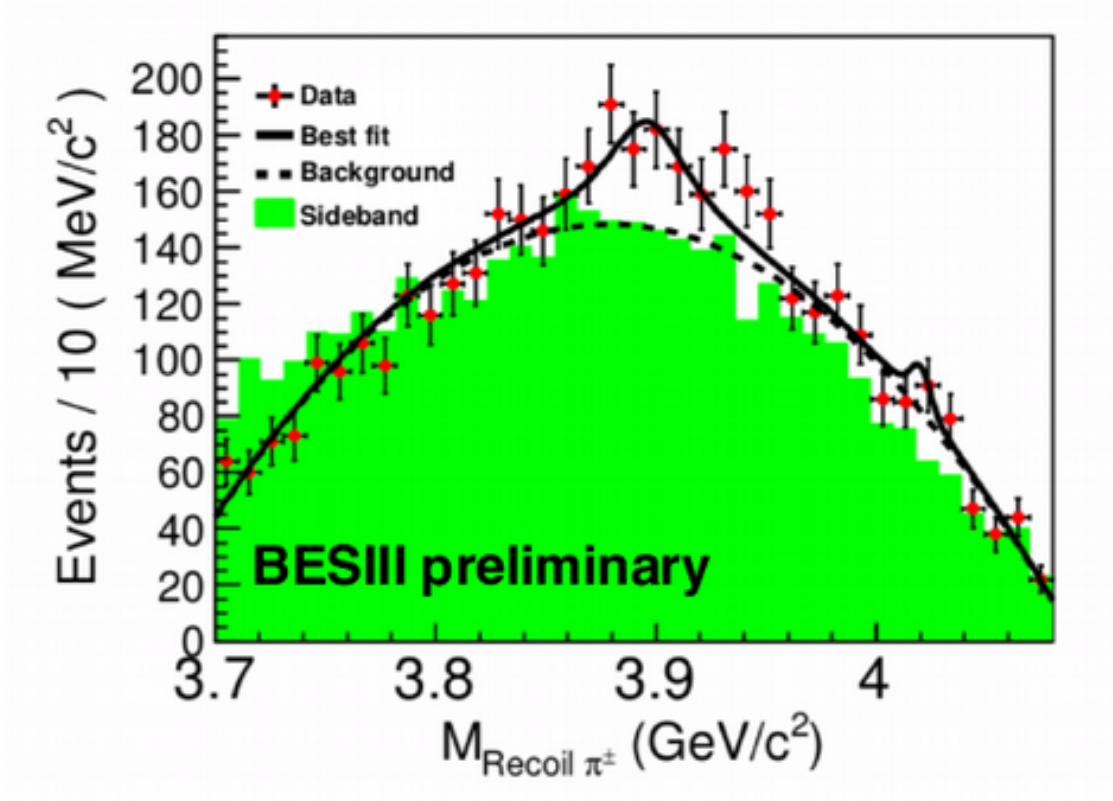}
\caption{Preliminary result- Recoil mass against $\pi^\pm$ in $e^+e^- \rightarrow \pi \rho \eta_c$. An evidence of $Z_c(3900)$ can be seen.}\label{fig:figure4.pdf}
\end{figure}

The charged charmonium-like $Z_c(3900)$ and $Z^{'}_c(4020)$ states are the most-striking evidence of exotics above $D\bar{D}^*$ threshold with a structure that has to contain at least four valence quarks. 
Whether these four quarks are arranged as compact tetraquarks or broad molecular states is not yet defined. Among the different models, the one proposed in Ref.~\cite{zc_models} suggests that it is possible to discriminate between the two hypotheses by studying the ratio between the $\rho \eta_c$ and the $\pi J/\psi$ and the $\pi h_c$ branching ratios for $Z_c(3900)$ and $Z^{'}(4020)$ respectively. Owing to the world largest data samples at 4.23, 4.26, and 4.36 GeV BESIII has reported a preliminary results of the evidence of $e^+e^- \rightarrow \pi Z_c$, $Z_c \rightarrow \rho \eta_c$ at 4.23 GeV as shown in Fig.~\ref{fig:figure4.pdf}. The process $e^+e^- \rightarrow \pi Z^{'}_c$, $Z^{'}_c \rightarrow \rho \eta_c$ is not seen in any of the dataset analysed.

\subsection{First measurement of $e^+e^- \rightarrow \pi D^0 \bar{D}^{*-}$ lineshape}
\begin{figure*}[t]
\centering
\includegraphics[width=140mm]{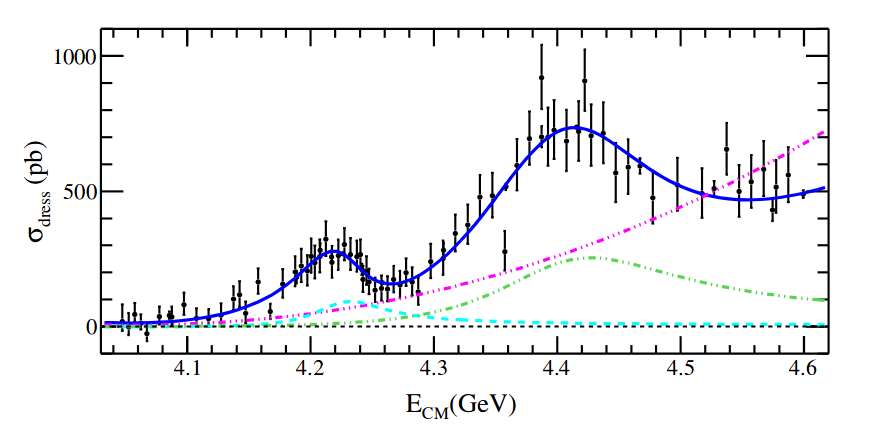}
\caption{Dressed cross section lineshape for the process $e^+e^- \rightarrow \pi D^0D^{*-}$ measured in 84 different energy values. A coherent sum of two Breit-Wigner plus a polynomial continuum is used to fit the lineshape.}\label{fig:figure5.pdf}
\end{figure*}

The possibility to directly access the $1^{--}$ charmonium-like states makes very easy for BESIII to collect large data samples to precisely determine the lineshape of different processes. In Ref.~\cite{besiii_piddstar} BESIII has reported the observation of two structures in $e^+e^- \rightarrow \pi^+D^0D^{*-}$ with a significance larger than 10$\sigma$ using 84 energy values collected from 4.05 and 4.6 GeV. The dressed cross-section is fitted with a coherent sum of two Breit-Wigner amplitudes plus a polynomial continuum as shown in Fig.~\ref{fig:figure5.pdf}. 

The parameters of the structure at higher mass vary largely with the parametrization used and a more detailed amplitude analysis is needed to provide a more robust description. The mass and the width of the low mass structure are instead more stable and are found to be $\mathrm{M(R)} = (4228.6 \pm 4.1 \pm 6.3)\, \mathrm{MeV}/c^2$ and $\Gamma(R) = (77.0 \pm 6.8 \pm 6.3)\, \mathrm{MeV}$. This is the first observation of a $Y(4220)$ state in an open charm final state. The mass found is compatible with the $Y(4220)$ observed in $\pi\pi h_c$ and $\pi \pi \psi(2S)$ final states, and slightly higher than the one found in $\omega \chi_{c0}$ and $\pi\pi J/\psi$.

\subsection{Updated measurement of $e^+e^- \rightarrow \omega \chi_{c0}(1P)$}
\begin{figure}[h]
\centering
\includegraphics[scale=0.2]
{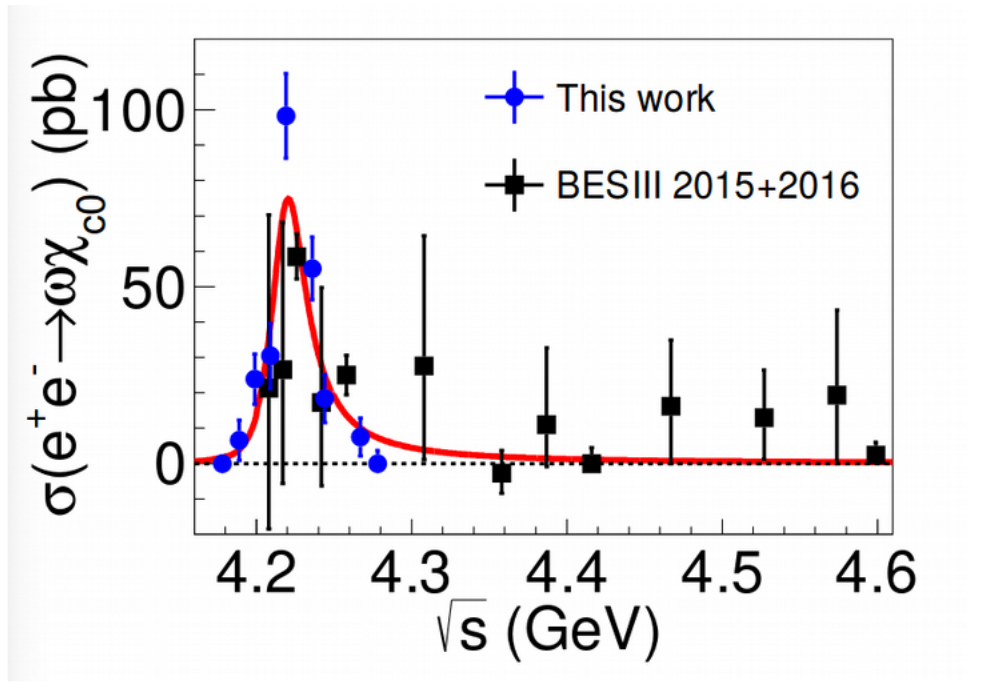}
\caption{Dressed cross section of the process $e^+e^- \rightarrow \omega \chi_{c0}$. Black squares are the measurement of \cite{old_omegachi_1, old_omegachi_2}, while the blue dots are the one of \cite{besiii_omegachi}.}\label{fig:figure6.pdf}
\end{figure}

BESIII has previously reported the cross section measurement of $e^+e^- \rightarrow \omega \chi_{c0}(1P)$ in Ref.~\cite{old_omegachi_1} and in Ref.~\cite{old_omegachi_2}. In 2017, a precise scan of the region between 4.178 and 4.278 GeV has been performed with 9 energy values collected with an integrated luminosity per energy set of roughly 500/pb. In Ref.~\cite{besiii_omegachi} BESIII has reported the updated measurements. The cross section is fitted with a single Breit-Wigner times a phase space factor that takes in account the energy spread as shown in Fig.~\ref{fig:figure6.pdf}. From the fit the mass $\mathrm{M} = (4218.5 \pm 1.6 \pm 4.0) \, \mathrm{MeV}/c^2$ and width $\mathrm{\Gamma} = (28.2 \pm 3.9 \pm 1.6)\, \mathrm{MeV}$ are extracted with a higher precision with respect to Refs.~\cite{old_omegachi_1, old_omegachi_2}. The observed parameters are not compatible with the $\psi(4160)$ ones, however, no conclusive indication on the nature of this state are possible with present experimental information.

\subsection{Study of $e^+e^- \rightarrow \pi^+ \pi^- D\bar{D}$}
Among the different proposal on the nature of the $Y(4220)$ the hypothesis of being a $D_1(2420)\bar{D}^*$ molecule can be tested since the model, presented in Ref.~\cite{Y_molecule}, give strong indications on the behaviour of the cross section. BESIII has recently completed the first study of the process $e^+e^- \rightarrow \pi^+ \pi^- D\bar{D}$ in Ref.~\cite{besiii_pipiDD}. This process can help test the molecular hypothesis.

\begin{figure*}[t]
\centering
\includegraphics[width=165mm]{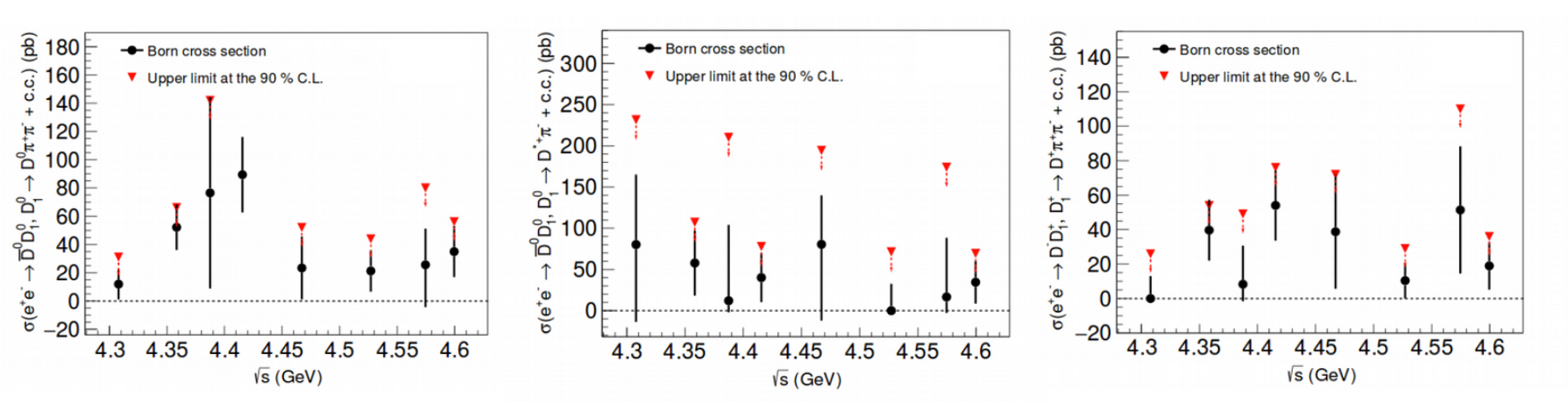}
\caption{Dressed cross section measurements for the processes $D_1(2420)^0 \rightarrow D^0 \pi^+ \pi^-$ (left),  $D_1(2420)^0  \rightarrow D^{*-} \pi$ (center), $D_1(2420)^+ \rightarrow D^+\pi^+\pi^-$ (right). Central values are calculated for all the center of mass energies and are shown as black dots. In those energy values in which there is no statistical significance to declare evidence, an upper limit is estimated and it is shown as a red arrow.}\label{fig:figure7.pdf}
\end{figure*}

BESIII has performed the search in 15 different center of mass energies between 4.09 and 4.6 GeV. Both neutral and charged $D$s were reconstructed with four and five  different tag modes respectively. To test the hypothesis it is necessary to search for the $D_1(2420)$ meson. In this work BESIII has chosen three decay modes: 1) $D_1(2420)^0 \rightarrow D^0 \pi^+ \pi^-$; 2) $D_1(2420)^0  \rightarrow D^{*-} \pi$; 3) $D_1(2420)^+ \rightarrow D^+\pi^+\pi^-$. The full lineshape is presented in Fig~\ref{fig:figure7.pdf}. The dark points are the measurements and the red arrows represent the upper limits. 

For the full spectrum analysed, BESIII has reported the observation of the first type of process at 4.42 GeV, evidence for the third type of process at 4.36 and 4.42 GeV and no evidence for the second type of process. By studying simultaneously the three lineshape no sharp rise of the $D_1(2420)\bar{D}$ cross section is observed: thus the molecular interpretation of the $Y(4220)$ seems disfavoured by present BESIII data.

\section{Interpreting the $\mathbf{Y(4660)}$ state}
There is an increasing interest in the $Y(4660)$ state. First observed in $\pi \pi \psi(2S)$ invariant mass by BELLE thanks to ISR return from $\Upsilon(4S)$ \cite{belle_y4660_1} it was searched also in other $\pi \pi$ transitions, but there are no evidence of such state in $\pi \pi J/\psi$ invariant mass. 

A structure with similar mass and width was reported by BELLE in $\Lambda_c \bar{\Lambda}_c$ invariant mass \cite{belle_y4660_2}, and no evidence in other open charm final state was reported. 

By studying the production cross section there is a striking evidence of an above threshold behaviour, since the coupling by the charmed baryon seems to be 11 times larger than the one with hidden-charmed meson ($\sigma(Y(4660) \rightarrow \Lambda_c \bar{\Lambda}_c) \sim 0.55 \, \mathrm{nb}$ and $\sigma(Y(4660) \rightarrow \pi \pi \psi(2S)) \sim 0.04 \, \mathrm{nb}$). This is an unique case, since the coupling of $Y(4220)$ and $Y(4360)$ states with baryons is very small. There are a few explanations for this evidence: the $Y(4660)$ state has no difference with respect to the other $Y$ states and another transition to an hidden-charm meson it exists and it can match the present difference; the $Y(4660)$ is a $\psi(2S)f_0(980)$ molecule, as suggested by Ref.~\cite{molecule_Y4660}; the $Y(4660)$ is an hidden-charm baryonium as proposed by Rossi-Veneziano in Ref.~\cite{baryonium_1} and more recently by Ref.~\cite{baryonium_2}. The hypothesis of the hadronic molecule seems slightly disfavoured by the fact that additional open charm decays are missing.

BESIII has already reported four cross section measurement of $e^+e^- \rightarrow \Lambda_c \bar{\Lambda}_c$ just above the production threshold \cite{besiii_lambdac}. The four measurements are in good agreement with BELLE data, but the trend seems different as pointed out by Ref.~\cite{meissner_lambdac}. The present BESIII data seems to reproduce what was observed in $e^+e^- \rightarrow p \bar{p}$ by CMD3 \cite{cmd3_ppbar}, where a step is observed at cross section opening. 

BEPCII has already completed an upgrade that will push the limit to the center of mass energy up to 4.7 GeV: this will guarantee the possibility to access the full $Y(4660)$ lineshape and test the different phenomenological predictions. BESIII will collect data in the region in the near future and its measurements will shed new light on the nature of the $Y(4660)$ thanks to high efficiency of measuring this processes close to thresholds. At the end, only two possibilities remain: either the two set of measurements agree, and in a not so far future the $Y(4660)$ parameters will be measured with an unprecedented precision; or the trend will remain different and it will be very complicated to sustain the hidden-charm baryonium models.

\section{Summary}
As experimentalists, our goal is to find new and deeper connections between these states in order to shed light on the nature of these states.

With the planned upgrade program, that involves also a new inner tracking system, an increased center of mass energy, and continuous injection BESIII will play a central role also in addressing the $Y(4660)$ nature.

The $XYZ$ states are one of the most exciting experimental observations in recent years in hadron spectroscopy. With the increasing precision of the new measurements, soon a much clearer picture on their nature will be unveiled. BESIII plays an unique role in this searches thanks to the possibility to directly access to vector charmonium-like states and accumulate large data sample: BESIII has already collected in 2019 additional 10 energies between 4.28 and 4.4 GeV to improve the knowledge on the $Y(4360)/Y(4320)$ state.


%

\bigskip 

\begin{thebibliography}{99}   
\bibitem{belle_x3872} S.-K.~Choi \textit{et al.} (BELLE Collaboration), Phys.~Rev.~Lett. {\bf 91}, 262001 (2003)

\bibitem{olsen} S.L.~Olsen, T.~Skwarnicki, D.~Zieminska, arXiv: 1708.040112v1 [hep-ph]

\bibitem{Voloshin} M.~Voloshin, \textit{Deciphering the XYZ states}, contribution in these proceedings PSN TueB1520

\bibitem{BESIII_nim} M.~Ablikim \textit{et al.} (BESIII Collaboration), Nucl.~Instr.~Meth.~Phys. A{\bf 614}, (2010) 345-399

\bibitem{besiii_omegaJpsi} M.~Ablikim \textit{et al.} (BESIII Collaboration), Phys.~Rev.~Lett~{\bf 122} 232002 (2019) arXiv: 1903.04695

\bibitem{babar_omegaJpsi} P.~del Amo~Sanchez \textit{et al.} (BABAR  Collaboration), Phys.~Rev. D{\bf 82} 011101 (2010)

\bibitem{gg_omegaJpsi} S.~Uehara \textit{et al.} (BELLE Collaboration), Phys.~Rev.~Lett. {\bf 104}, 092001 (2010)

\bibitem{b_to_omegaJpsi} S.-K.~Choi \textit{et al.} (BELLE Collaboration), Phys.~Rev.~Lett. {\bf 94}, 182002 (2005)

\bibitem{x3872_molecule} F.~K.~Guo, C.~Hanhart, U.-G.~Meissner, Q.~Wang and Q.~Zhao, Phys.~Lett.~B{\bf 725}, 127 (2013)

\bibitem{voloshin_pi0} S.~Dubynskiy, M.~Voloshin, Phys.~Rev.~D{\bf 77}, 014013 (2008)

\bibitem{BESIII_pi0} M.~Ablikim \textit{et al.} (BESIII Collaboration), arXiv: 1901.03992

\bibitem{zc_models} A.~Esposito, A.~L.~Guerrieri, A.~Pilloni, Phys.~Lett.~B{\bf 746}, 194 (2015)

\bibitem{besiii_piddstar} M.~Ablikim \textit{et al.} (BESIII Collaboration), Phys.~Rev.~Lett. {\bf 122}, 102202 (2019)

\bibitem{old_omegachi_1} M.~Ablikim \textit{et al.} (BESIII Collaboration), Phys.~Rev.~Lett.~ {\bf 114}, 092003 (2015)

\bibitem{old_omegachi_2} M.~Ablikim \textit{et al.} (BESIII Collaboration), Phys.~Rev.~D{\bf 93}, 011102 (2016)

\bibitem{besiii_omegachi} M.~Ablikim \textit{et al.} (BESIII Collaboration), Phys.~Rev.D{\bf 99} 091103 (2019) 

\bibitem{Y_molecule} M.~Cleven \textit{et al.}, Phys.~Rev.~D{\bf 90}, 074039 (2014)

\bibitem{besiii_pipiDD} M.~Ablikim \textit{et al.} (BESIII Collaboration), arXiv: 1903.08126


\bibitem{belle_y4660_1} X.~L. Wang \textit{et al.} (BELLE Collaboration), Phys.~Rev.~D{\bf 91}, 112007 (2015)

\bibitem{belle_y4660_2} G.~Pakhlova \textit{et al.} (BELLE Collaboration), Phys.~Rev.~Lett. {\bf 101}, 172001 (2008)

\bibitem{molecule_Y4660} F.-K.~Guo \textit{et al.}, Phys.~Rev.~D{\bf 82}, 094008 (2010)

\bibitem{baryonium_1} G.~C.~Rossi, G.~Veneziano, Nucl.~Phys.~B{\bf 123}, 507 (1977)

\bibitem{baryonium_2} G.~Cotugno, R.~Faccini, A.~D.~Polosa and C.~Sabelli, Phys.~Rev.~Lett. {\bf 104}, 132005 (2010)

\bibitem{besiii_lambdac} M.~Ablikim \textit{et al.} (BESIII Collaboration), Phys.~Rev.~Lett.~{\bf 120}, 132001 (2018)

\bibitem{meissner_lambdac} L.-Y.~Dai, J.~Haidenbauer, U.-G.~Meissner, Phys.~Rev.~D{\bf 96}, 116001 (2017)

\bibitem{cmd3_ppbar} R.~R.~Akhmetshin \textit{et al.} (CMD-3 Collaboration), arXiv: 1808.00145

\end{thebibliography}

\end{document}